\documentclass[epj]{svjour}
\usepackage{graphics}
\usepackage{epsf}
\begin{document}
%
\title{Conserved Dynamics and Interface Roughening in 
Spontaneous Imbibition : A Phase Field Model
}
\author{M.\ Dub\a'e\inst{1,2} \thanks{\emph{Present address: 
Center for the Physics of
Materials, McGill University, 3600 rue University, Montr\a'eal, Qu\a'ebec,
Canada H3A 2T8   }   }, M.\ Rost\inst{1,2}, 
K.\ R.\ Elder\inst{3}, M. Alava\inst{2}, S.\ Majaniemi\inst{1,2} 
\and T.\ Ala-Nissila\inst{1,2,4}
}                     
%
%
\institute{Helsinki Institute of Physics, P.O. Box 9, FIN--00014,
University of Helsinki, Helsinki, Finland \and 
Laboratory of Physics, P.O. Box 1100, Helsinki University of Technology, 
FIN--02015 HUT, Espoo, Finland \and 
Department of Physics, Oakland University,
Rochester, MI 48309-4487, U.S.A. \and
Department of Physics, Brown University, Providence, RI
02912--1843, U.S.A.
}
\authorrunning{Dub\a'e et al.}
\titlerunning{Conserved Dynamics in Imbibition}
%
%
\abstract{The propagation and roughening of a fluid-gas interface through a
disordered medium in the case of capillary driven spontaneous
imbibition is considered. The system is described by a conserved
(model B) phase-field model, with the structure of the disordered
medium appearing as a quenched random field $\alpha({\bf x})$. The
flow of liquid into the medium is obtained by imposing a
non-equilibrium boundary condition on the chemical potential, which
reproduces Washburn's equation $H \sim t^{1/2}$ for the slowing
down motion of the average interface position $H$. The interface is
found to be superrough, with global roughness exponent $\chi \approx
1.25$, indicating anomalous scaling. The spatial extent of the
roughness is determined by a length scale $\xi_{\times} \sim
H^{1/2}$ arising from the conservation law. The interface 
advances by avalanche motion, which causes temporal multiscaling and
qualitatively reproduces the experimental results
of Horv\a'ath and Stanley [Phys.\ Rev.\ E {\bf 52} 5166 (1995)] on the
temporal scaling of the interface. 
} 
\maketitle

\section{Introduction}

The dynamics of driven interfaces in disordered media is a subject of 
intense interest in nonequilibrium statistical mechanics. It is well
established that for sufficiently strong driving, the interface feels
an effective smeared out ``thermal'' noise and its fluctuations present 
all the typical phenomena of scale invariance of driven 
systems \cite{Krug_97}.
In the opposite case of weak driving, the quenched nature of the noise
becomes apparent and the interface may reach a pinned state,
characterised by completely different scaling exponents \cite{Kardar_98}.

An apparently easy experiment to perform is to monitor the motion of
an invading liquid front in a porous medium. Many experiments have
been done with Hele-Shaw cells
\cite{Rubio_89,Horvath_91,He_92,Delker_96,Dougherty_98}, and the
spontaneous imbibition of water in paper has also been considered
\cite{Buldyrev_92,Barabasi_92,Family_92,Buldyrev_ph,Amaral_94,Horvath_95,Kumar_96,Kwon_96,Nagamine_96,Zik_97}.
In this last case, capillary forces arising from the porous structure
drive the liquid until loss of water by evaporation or hydrostatic pressure
balance the driving. The eventual pinning of the front has received a
lot of attention \cite{Buldyrev_92,Barabasi_92,Amaral_94,Kumar_96},
and some experiments also examined the complete dynamical process
\cite{Family_92,Horvath_95,Kwon_96,Zik_97}. It is generally believed
that phenomena in the critical region of the depinning transition can
be described by local theories, i.e., the physics is governed by 
an equation that couples the interface locally with itself
and the quenched randomness  \cite{Kardar_98}. 
The 
spatial configurations of a pinned imbibition front seem to exhibit scaling
properties well described by the ``Directed Percolation Depinning''
(DPD, or Quenched Kardar-Parisi-Zhang, QKPZ) universality class
\cite{Buldyrev_92,Barabasi_92,Amaral_94,remark_poor_scaling}. An
intuitively motivated lattice model of DPD compares well to
experimental findings on the stopped front
\cite{Buldyrev_92,Barabasi_92}, and a modified version of the model
addresses the influence of evaporation on the scaling properties of
the pinned front \cite{Amaral_94}.

However, these models neglect the fact that liquid has to be transported
through the medium in order to drive the front, a nontrivial phenomenon
in itself \cite{Bouchaud_93,Scheidegger_57,Ganesan_98}. For example,
viscous fluid transport explains why the invading front continuously
slows down even {\em without} evaporation or gravity, a result that 
has been well
established in the literature \cite{Washburn_21,experiments,Dube_Exp}. 
In local models this has to be put in
rather artificially \cite{Amaral_94}. The temporal correlations of the
fluctuations should also reflect this nonlocality. It is thus doubtful
whether any local model can explain the experimental results of
Horv\'ath and Stanley, focusing on the dynamical scaling of the
interface \cite{Horvath_95}.

The main concern in this paper is to analyse in detail a simple model
of a propagating liquid-gas interface in a disordered medium, already
introduced in Ref.\ \cite{dube_99}. The previous companion paper
\cite{Dube_Exp} presents a general overview of the experimental and
macroscopic aspects of imbibition and discusses in particular the role
of the fluid conservation law. The main goal in building the present 
model is thus to incorporate local liquid conservation to the interfacial
dynamics. In Section \ref{modelpresent}, it is shown that this can be
achieved through a generalised Cahn-Hilliard equation. The model can
then be applied to two different physical situations: a freely rising
front, and a front propagating against a steady motion of the paper
towards the liquid reservoir, leading to stationary fluctuations. 
A central feature of the model is the existence of a length scale 
$\xi_{\times}$, emerging from the interplay  between interfacial
tension and liquid conservation. This reflects the inherently
non-local nature of the dynamics. The dynamical evolution of the model
is numerically integrated in Section \ref{numerics}, for the two
different setups. In Section \ref{analysis} further implications of
our modeling are discussed, and dynamical scaling in the experiments
of Horv\a'ath and Stanley \cite{Horvath_95} is reinterpreted. We
conclude with suggestions for future experimental work. 
The Appendices 
contain generalisations of the model to cases where gravity and/or 
evaporation might be important and a brief description of the steps
required to obtain the interface equation, Eq. (\ref{non-local}).

\section{Phase Field Model of Imbibition}
\label{modelpresent}

\subsection{Definition of the model}

A full model of the dynamics of a liquid invading a random medium
based on a microscopic treatment is a formidable problem. The dynamics
of the advancing interface between the liquid and the (dry) solid
should however be amenable to a discussion at the coarse grained level.
In this spirit, a phase-field model is used to represent the spatial
configuration of wet and dry ``phases''. The field of interest is a
locally conserved quantity $\phi ({\bf x},t)$, defined on the
half-plane $\{{\bf x} \! \equiv \! (x,y) \vert y \! \ge \! 0 \}$, with
values $\phi \! = \! +1 (-1)$ for the wet (dry) phase. A free energy
of the form $\mathcal{F} \{ \phi \} = \int d^dx \; [ (\nabla \phi)^2
\! /2 + V(\phi)]$ is chosen, with a double well potential $V(\phi)$ of
the standard Ginzburg-Landau form, to which is added a linear tilt,
\begin{equation}
\label{pf_potential}
V({\bf x},\phi({\bf x},t)) \equiv -\frac{1}{2} \; \phi^2({\bf x},t)
+\frac{1}{4}  \; \phi^4({\bf x},t) - \alpha({\bf x}) \phi({\bf x},t).
\end{equation}
The double well potential, together with the gradient term, ensures 
the existence of a well defined
interface, and the quenched random field $\alpha({\bf x})$ represents
the random (coarse grained) structure of the medium. The first and the
second moments associated with the underlying distribution of the
random medium are given by $\langle \alpha({\bf x}) \rangle \! = \!
\bar \alpha$, and $\langle \alpha({\bf x}) \alpha({\bf x'})\rangle \!
- \! \bar \alpha^2 \! = \! (\Delta \alpha)^2 \delta({\bf x} \! - \!
{\bf x'})$. It is thus assumed that $\alpha$ is spatially
uncorrelated, which may be a good approximation in the case of ordinary
paper (the areal mass density has only short-range correlations
\cite{Provatas_96}).

The dynamics of the conserved variable $\phi ({\bf
x},t)$ is determined by a continuity equation $\partial_t \phi +
\nabla \! \cdot \! {\bf j} = 0$, where the current ${\bf j} ({\bf
x},t) = - \nabla \mu ({\bf x},t)$ is related to the gradient of the
chemical potential $\mu ({\bf x},t)  \! \equiv \! - \delta \mathcal{F}
\! / \! \delta \phi({\bf x},t)$. The resulting equation of motion,
\begin{eqnarray}
\label{pf_eq}
\partial_t \phi({\bf x},t) & = & \nabla^2 \mu({\bf x},t) \\
& = & \nabla^2 \left[
- \phi({\bf x},t) + \phi^3({\bf x},t) - \nabla^2 \phi({\bf x},t) -
\alpha({\bf x})\right], \nonumber 
\end{eqnarray}
is essentially the Cahn-Hilliard equation \cite{Cahn_58}, also used to
study critical fluctuations and phase ordering in presence of a
conservation law (model B dynamics \cite{Hohenberg_77,Bray_94}). The
variable $\alpha ({\bf x})$ here plays the role of the local chemical
potential at the interface thus controlling the flux.

In model B dynamics \cite{Bray_94}, the domain walls are driven by the
difference between incoming and outgoing current ${\bf j} \! = \! -
\nabla \mu$. In the sharp interface limit, and for a slowly moving
front, $\mu ({\bf x},t)$ changes quasistatically, always satisfying
$\nabla^2 \mu \! = \! 0$ in the bulk, plus the appropriate boundary
conditions. At the interface, $\mu$ must obey the Gibbs-Thomson
boundary condition
\begin{equation}
\label{gibbsthompson}
\Delta \phi \; \mu \vert_{int} = \Delta V - \sigma \mathcal{K},
\end{equation}
where $\mathcal{K}$ is the curvature, $\sigma = 2 \sqrt{2}/3$ is the
surface tension of the model, the miscibility gap $\Delta \phi =
\phi_+ - \phi_-$ and  $\Delta V = V(\phi_+)-V(\phi_-)$. The quantities
$\phi_{\pm}$  are the {\it equilibrium} values of the phase field,
defined by the usual tangent construction \cite{Bray_94,Langer_92}.
The interface motion is then determined by the normal velocity $v_n =
- \partial_n \mu \vert_-^+$.

\subsection{Freely rising and stationary fronts}

\begin{figure}
\epsfxsize=3.2in \epsfysize=3.2in
\epsfbox{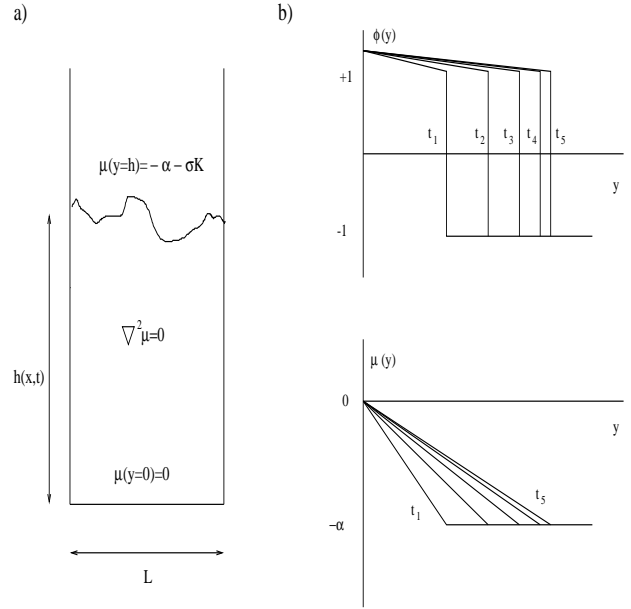}
\caption{Setup of the imbibition model. The system is defined on the
plane $y>0$, with lateral extent $L$. The average position of the
interface is represented by $H(t)$. The chemical potential obeys
Laplace's Equation in the bulk $\nabla^2 \mu =0$ with
Gibbs-Thomson boundary condition $\mu = -\alpha({\bf x}) - \sigma
 \partial^2_x h(x,t)$ at the interface and the imposed value
$\mu(y=0)=\alpha_0 = 0$ at the bottom.
}
\label{cartoon}
\end{figure}

With appropriate boundary conditions, the model presented above can
encompass the typical experimental setups of imbibition, as depicted
in Fig.\ \ref{cartoon}. These are such that the value of the chemical
potential $\mu (x,y=0) = \alpha_0 \neq \bar{\alpha}$ is {\em imposed}
at the bottom end  while the top end of the system is kept dry 
(i.e., $\partial_y \mu(y \to \infty) = 0$ and 
$\phi(y \! \to \! \infty) = -1$). 
This concept can easily be explained
by a simplified situation where the quenched field $\alpha ({\bf x})$
is homogeneous and equal to a constant $\bar{\alpha}$. In this case,
an equilibrium interface would be obtained by letting $\mu =
-\bar{\alpha}$ throughout the whole system, with $\phi_{\pm} = \pm
1$. On the other hand, imposing the boundary condition $\mu (x, y=0) =
\alpha_0$ creates an imbalance in the chemical potential causing the
interface to advance. If the interface at time $t$ is at a height
$H(t)$, the chemical potential, as given by the Laplace equation, is
\begin{eqnarray}
\mu ({\bf x},t) = \mu(y,t) & = & \alpha_0 - (\bar{\alpha} - \alpha_0) 
\frac{y}{H(t)}, \mbox{ for $y \leq H$}; \nonumber \\ 
\mu(y,t) & = & - \bar \alpha, \mbox{ for $y > H$},
\end{eqnarray}
resulting in a time evolution \footnote{This is valid only if
$\bar{\alpha} - \alpha_0 \ll 1$. A more complete expression is given
in Appendix \ref{real}.} 
\begin{equation}
\label{washburn}
\frac{ d H(t)}{dt} = \frac{\bar{\alpha}-\alpha_0}{2 H(t)}.
\end{equation}
Thus, the further the interface is from the reservoir, the smaller 
its velocity. This classical result, known as the Washburn equation, is
well established experimentally, although discrepancies may arise
\cite{Washburn_21,experiments,Dube_Exp,Zik_98}.

A slight modification to the model can be used to reproduce the
experimental setup of Horv\'ath and Stanley \cite{Horvath_95}.  In
their experiment, the interface is forced to stay at a fixed mean
height $H$ by constantly pulling down the paper towards the reservoir
of liquid. Within the phase
field model it is easy to add a constant downward drift ${\bf v} = - v
\hat {\bf y}$, so that
\begin{eqnarray}
\label{horvath_eq}
\partial_t \phi({\bf x},t) + {\bf v} \! \cdot \! \nabla
\phi ({\bf x},t)
= \nonumber \\  
\nabla^2 \left[ - \phi({\bf x},t) + \phi^3({\bf x},t)
 - \nabla^2 \phi({\bf x},t) - \alpha({\bf x \! - \! v}t)
\right],
\end{eqnarray}
keeping the same boundary conditions as for the freely rising
column. Thus the interface between the wet and the dry region is kept
at a height $H$ where a rising interface would have a velocity $-{\bf
v}$, or
\begin{equation}
H \! = \! \frac{\bar{\alpha} - \alpha_0}{2 v}.
\label{fixed-h}
\end{equation}

\subsection{Equation of motion and correlation length}

\begin{figure}
\epsfxsize=3.2in \epsfysize=3.2in
\epsfbox{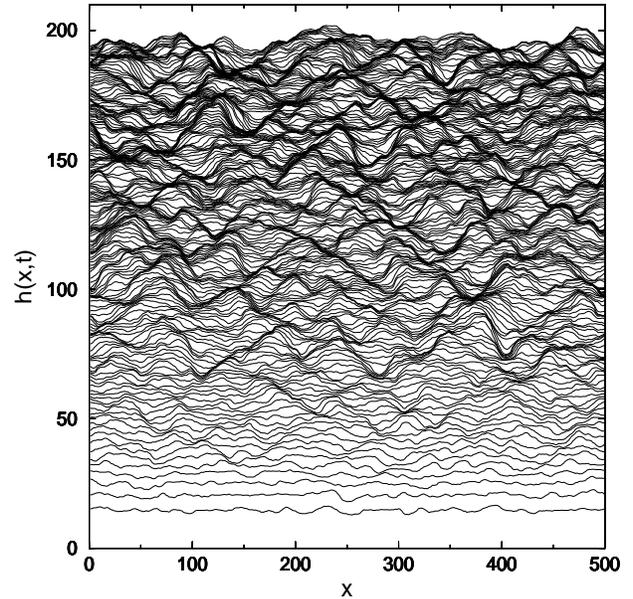}
\caption{Front configurations of a rising interface at equal time
intervals $\Delta t=10^3$. Their average separation becomes smaller
as the 
front slows down due to the conservation law.
}
\label{fronts_fig}
\end{figure}

In presence of quenched disorder, via the field $\alpha ({\bf x})$,
the interface will start to roughen, as shown in Fig.\
\ref{fronts_fig}. An immediately noticeable feature is that the
interface looks extremely rough locally but appears
smooth on large length scales.
This should indeed be expected intuitively, since the physics of
the phenomenon is such that parts of the interface
ahead of the average interface position
have a smaller instantaneous {\em local} velocity. They
are thus eventually ``caught up'' by the average interface. Likewise,
retarded parts of the interface tend to catch up with the average
interface position.  This idea is indeed confirmed by the numerical
results, and can further be used to define the spatial range over
which the correlated roughness may be seen.

A key step in understanding the physics consists in writing an
interface equation for the present model. A single-valued one
dimensional interface $y=h(x,t)$ is assumed, and the Green's function
of the problem is defined through the relation
\begin{equation}
\label{gr_def}
\nabla^2  G(x,y|x',y') = - \; \delta (x-x') \; \delta(y-y'),
\end{equation}
for the range $-\infty < x,x' < \infty$, $0 < y,y' < \infty$, with
Dirichlet boundary conditions. The half-plane must be used, since the
presence of an ``infinite reservoir'' at position $y=0$ breaks the
translational symmetry in $y$. Without any loss of generality,
$\alpha_0=0$ is set from now on. The standard procedure
\cite{Bray_94,Langer_77,Kawasaki_82}, exposed in Appendix \ref{proj}, 
may then be followed to obtain
the integro-differential equation of motion
\begin{eqnarray}
\int_{-\infty}^{\infty} dx' \; G (x,h(x,t)|x',h(x',t)) \;
\frac{\partial h(x',t)}{\partial t} = \nonumber \\ 
\eta (x,h(x,t)) + \sigma {\cal K}
\label{non-local}
\end{eqnarray}
with the half-plane Green's function
\begin{equation}
\label{gr_form}
G(x,y|x',y') = \frac{1}{4\pi} \ln \; \frac{(x-x')^2 +
(y-y')^2}{(x-x')^2+(y+y')^2}.
\end{equation}
The quenched noise is written as $\eta(x,h) \equiv \int dy
\phi_{0}'(y-h(x,t)) \, \alpha(x,y) \sim 2 \alpha (x,h)$ in the sharp
interface limit. The Gibbs-Thomson boundary condition, $\mu|_{int}
\sim {\cal K}$ can be immediately obtained from Eq.\ (\ref{non-local})
in the limit $\dot{h}=0$ since $\eta$ is the chemical potential at the
interface. Analogous non-local equations will arise in the context of
directional solidification, pattern selection in Laplacian fluid flow
\cite{Krug_91} and step growth \cite{Bales_90}. The novel features
here are the broken translational invariance and the presence of
quenched noise. The interface fluctuations are thus intimately coupled
to both the average position and the average velocity of the
interface, a result that comes out self-consistently from the
model. This is quite different from local types of equations or
models. It should be particularly noted that the presence of a
conservation law does not result in a ``conserved'' interfacial
equation. Likewise, nonlinear equations with long range kernels
\cite{Mukherji_97} do not apply to the situation encountered here.

The difference between local models and Eq.\ (\ref{non-local}) becomes
even clearer if the interface is linearised in small deviations around
the mean interface position $H(t)$ to obtain
\begin{eqnarray}
\label{effint}
\dot{h_k} \left( 1 - e^{-2 \vert k \vert H} \right)
+  |k| \dot H \; h_k \left( 1 + e^{-2 \vert k \vert H}
\right) = \nonumber \\ 
|k| \left( \{ \eta (t)\}_k -  \sigma k^2 h_k \right),
\end{eqnarray}
where $h_k$ are the Fourier components of $h$ and $H=h_0$ is the
average interface position. Note that the interface configuration
enters the disorder term in Eq.\ (\ref{effint}) in a fundamentally
nonlinear way,
\begin{equation}
\{ \eta (t)\}_k \equiv \int_{x} e^{-ikx} \eta(x,h(x,t)).
\end{equation}

This equation already yields important information for the roughening
process. For example, the limit $k \rightarrow 0$ reproduces the
slowing down of the front as given by Eq.\ (\ref{washburn}). It also
reveals the different length scales in the problem. The average height
of the interface separates two regimes of lateral scale. For $kH \ll
1$, Eq.\ (\ref{effint}) becomes
\begin{equation}
\frac{dh_k}{dt} + \frac{1}{H} \frac{dH}{dt} h_k + \frac{\sigma}{2H}
k^2 h_k = \frac{1}{2H} \{ \eta (t)\}_k,
\label{eff_local}
\end{equation}
and in the opposite short scale limit, $kH \gg 1$
\begin{equation}
\frac{dh_k}{dt} + \vert k \vert \; \frac{dH}{dt} h_k + \sigma
\vert k \vert^3 h_k =  k \{ \eta (t)\}_k.
\label{eff_nonlocal}
\end{equation}

The average interface height $H$ is thus also a lateral length scale.
If two points are separated by a distance $r \gg H$, they are not
connected through the bulk of the system and receive liquid from the
reservoir independently. In that sense, the dynamics of the interface
on larger scales is {\em local}, although the slowing down of the
interface is inherently a non-local phenomenon, reflecting liquid
transport through the medium. In the opposite limit of $r \ll H$, the
two interface points will be coupled through the bulk and compete for
liquid coming from the same region behind the front. The dynamics of
the interface then becomes fully {\em non-local}. 

However, in both limits the damping terms induce a separating length
scale $\xi_\times = (\sigma/\dot H)^{1/2} = (\sigma H/\bar
\alpha)^{1/2}$. For $\xi_\times k \gg 1$ the fluctuations of the
interface are damped due to the line tension $\sigma$, while for
$\xi_\times k \ll 1$, it is due to flow from the reservoir. By this
mechanism, it is expected that the front is smoother on length scales
larger than $\xi_{\times}$ as compared to smaller scales.

This length scale is closely related to the Mul\-lins-Se\-ker\-ka
instability of driven Laplacian fronts \cite{Mullins_64}, although the
situation is reversed here. Because fluid is transported towards the
front from behind, advanced (retarded) parts of the interface receive
less (more) mass than the average and the front is {\it stabilised} at
long length scales. This result can intuitively be understood as
follows. Due to the Gibbs-Thomson effect a local ``bulge'' of vertical
extent $W$ and lateral size $\xi$ alters the chemical potential by
$\Delta \mu \simeq \sigma W/\xi^2$. On the other hand, the average
gradient in $\mu$ in the bulk liquid induces a difference $\Delta \mu
\simeq \bar \alpha W /H$ across a vertical distance $W$. These two
differences balance each other at a length given by
\begin{equation}
\label{xi_cross}
\xi_\times \; \simeq \; \sqrt{\sigma H/\bar \alpha}.
\end{equation}
The length scale $\xi_{\times}$ is a static quantity, but in a rising
liquid column it becomes dynamical through the time dependence of $H(t)$,
i.e., $\xi_{\times} = \xi_{\times}(t) \sim (H(t))^{1/2} \sim t^{1/4}$.
It must also be noted that it is {\em not} a {\em truly} dynamical 
correlation length in the sense of kinetic roughening theories \cite{Krug_97}. 
However, it is a time dependent upper cutoff for correlated
fluctuations increasing with time, and therefore can be interpreted as
a dynamic correlation length.

To draw any further analytical conclusions from Eq.\ (\ref{effint}) is
extremely problematic due to the difficulties encountered with the
quenched noise $\{ \eta \}_k$ (which are analysed e.g.\ in
\cite{Leschhorn_QEW}). Furthermore, Eq.\ (\ref{effint}) is a linear
approximation to Eq.\ (\ref{non-local}). Although the length scales
come out correctly, it cannot be expected a priori that the correct
scaling properties of the interface will be obtained. An analysis
along the lines of Ref.\ \cite{Leschhorn_QEW} may prove insufficient,
and simplified treatments, such as those attempted in Ref.\
\cite{Ganesan_98} are inadequate.

\section{Numerical Analysis}
\label{numerics}

The interface fluctuations in the presence of quenched disorder were
analysed by numerical integrations of Eq.\ (\ref{pf_eq}) and Eq.\
(\ref{horvath_eq}). The position of the interface $h(x,t)$ at each $x$
was defined by the zero of the phase field, i.e.\ by
$\phi(x,h(x,t))=0$ determined by linear interpolation between the
points of the numerical grid. Overhangs, appearing for strong disorder
but otherwise absent were ignored by taking the lowest or highest zero
of $\phi$ above a given foot point $x$. No quantitative differences
were seen between these two choices.

The disorder $\alpha({\bf x})$ is an independently distributed random
variable on each grid point, with mean $\bar{\alpha}$, standard
deviation $\Delta \alpha$ and chosen from different types of
distributions (gaussian, uniform on a finite interval, and
exponential). Without loss of generality, the lower boundary condition
is chosen such that $\mu (x,y \! = \! 0)=0$, leading to $\phi(x,y \! =
\! 0) = \phi_0$, with $\phi_0$ the solution of $-\phi_0+\phi_0^3
= \bar{\alpha}$.

To evaluate any scaling behaviour the first quantities of interest are
the total width of the front,
\begin{equation}
W^2 (t) =  \langle \overline{(h(x,t)-H(t))^2} \rangle,
\end{equation}
and the related spatial two-point correlation functions of the $q$th
moments
\begin{equation}
G_q (r,t) =  \langle \overline{| h(x+r,t)-h(x,t) |^q} \rangle^{1/q}.
\end{equation}
The case $q \! = \! 2$ is directly related to the structure factor
$S(k,t) = \langle \overline{h_k (t) h_{-k} (t)} \rangle$. 
In the above equations
the brackets denote an average over different realisations of
$\alpha$, and the overbar a spatial average over the system. In
presence of a stationary state also temporal averages can be taken.

The standard Family-Vicsek scaling assumption rests on a dynamical
correlation length $\xi_t \sim t^{1/z}$ where $z$ is the dynamical
exponent, related to the decay of fluctuations along the interface.
The maximal value it can attain is the system size $L$ at which point
the interface is said to be in a ``saturated'' stationary state. The
two-point correlation function then has a scaling form $G_2 (r,t) =
r^{\chi} f(r/\xi_t)$ where $f(u) = const.$ for $u \ll 1$ and $f(u) \sim
u^{-\chi}$ for $u \gg 1$, a form which introduces the roughness
exponent $\chi$ and defines the associated growth exponent $\beta =
\chi/z$. The structure factor has a corresponding scaling form, $S(k,t)
= s(k \xi_t) / k^{1+2\chi}$ with the scaling function $s (u)$
constant for $u \gg 1$ and $s(u) \sim u^{1+2\chi}$ for $ u \ll 1$. 

This picture may turn out to be incomplete or even wrong for the
following reasons. First, the structure factor may contain an explicit
time dependence besides $\xi_t$, $S(k,t) \sim t^{2 \kappa} s(k \xi_t)
/k^{1+2 \chi}$, which is sometimes referred to as intrinsic
anomalous scaling \cite{Lopez_97}. Second, if the interface is
superrough, a case characterised by $\chi > 1$, then $G_2(r,t) \sim
\xi_t^\chi (r/\xi_t)^{\chi_{\rm loc}}$ with a local exponent
$\chi_{\rm loc} = 1$, since, by construction, $G_2(r)$ cannot increase
faster than $r$ \cite{Leschhorn_comment}. In contrast to the standard
Family-Vicsek picture, in both these cases the correlation function for
$r < \xi_t$ do not saturate as long as $\xi_t$ increases
\cite{Krug_97}. This can be parametrised by the scaling of the local
slopes $G_2 (r=1,t) \sim t^{(\chi-\chi_{\rm loc})/z}$. Third, the lateral
$\xi_t$ and the vertical scale $W$ may not be enough to characterise
the interface fluctuations, and different moments of $G_q$ may possess
different scaling exponents $G_q(r) \sim r^{\chi_q}$ (see e.g.\
\cite{Krug_94}).

The interface scaling behaviour may also be observed in the temporal 
correlation functions
\begin{equation}
C_q(t) = \langle \overline{ \vert h(x,t+s) - H(t+s) - h(x,s) + H(s)
\vert^q }\rangle^{1/q},
\end{equation}
which increase as $C_q (t) \sim t^{\beta_q}$ at short time differences
$t$. Of course, this definition makes only sense in a steady state,
under time-translational invariance. It therefore applies to the
analysis of Eq.\ (\ref{horvath_eq}), where the average interface
height is kept fixed by pulling down the paper at constant velocity
$v$.

\subsection{Freely rising fronts}
\label{freecolumn}

In this Subsection, the classic imbibition experiment is considered. A
liquid front is allowed to invade the porous medium starting from a
reservoir placed at $y=0$. Successive configurations obtained from 
numerical integration of Eq.\ (\ref{pf_eq}) are presented in Fig.\
\ref{fronts_fig}. The time difference between the curves
is constant ($\Delta t = 10^3$), and the slowing down of
the interface positions becomes apparent 
from the fact that they lie closer together
the higher the front gets.

\begin{figure}
\epsfxsize=3.2in \epsfysize=3.2in
\epsfbox{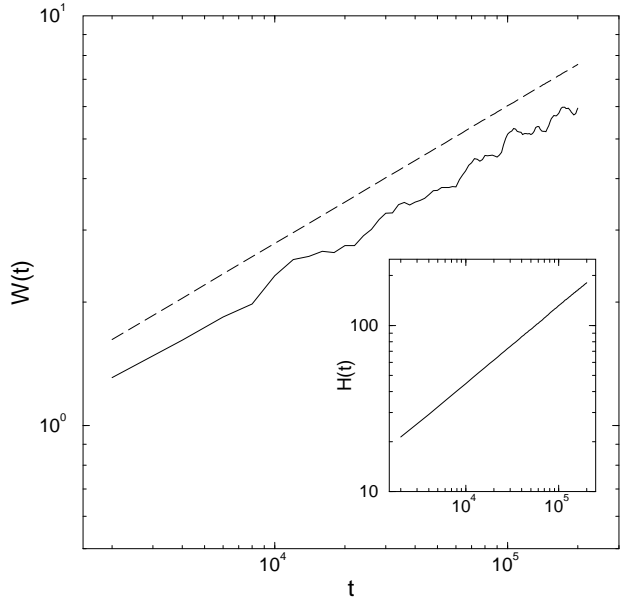}
\caption{Increase of the width of the interface as a function of time.
The data are for a system of lateral extent $L = 256$, with
$\bar{\alpha} = 0.2$ and $\Delta \alpha = 0.2$. The straight dashed
line has slope $\beta = 0.32$. The inset shows the average position of
the interface $H^2 (t) = \bar{\alpha} t$, in agreement with Eq.\
(\ref{washburn}).
}
\label{W-t}
\end{figure}

In the presence of quenched disorder $\alpha ({\bf x})$ the total width of
the interface increases as a power of time. Fig.\ \ref{W-t} shows that
$W(t) \sim t^{\beta}$ with $\beta \approx 0.3$. In the same graph the
progression of the average interface height $H$ is seen to follow
Washburn's behaviour as expected from the analytic arguments.

\begin{figure}
\epsfxsize=3.2in \epsfysize=3.2in
\epsfbox{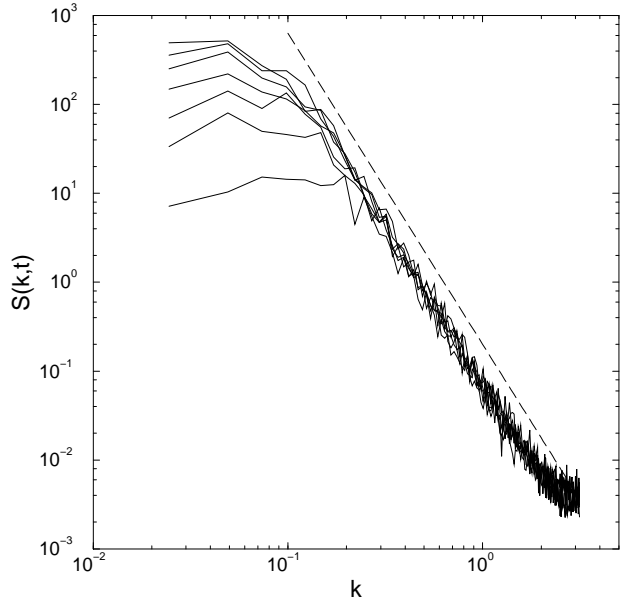}
\caption{Structure factor for a system with $L = 256$, $\bar{\alpha}
= \Delta \alpha =0.2$ at different times. The data are for times $t=5
\times 10^3$ (lowest curve) to $t=5 \times 10^4$ (upper curve) at
intervals of $10^4$. The dashed line has a slope $-3.5$ indicating a
global roughness exponent $\chi \sim 1.25$. A weak intrinsic anomalous
scaling, in the sense of Ref.\ \cite{Lopez_97} may be present, but
cannot be clearly identified.
}
\label{strf_fr}
\end{figure}

It turns out to be impossible to determine the global roughness exponent
$\chi$ from its definition in terms of the dependence of the saturated
width on the lateral system size. Instead, the structure factor $S(k,t)$
is used. It is plotted on Fig.\ \ref{strf_fr} for a system of size $L =
256$, with $\bar{\alpha} = 0.2$ and $\Delta \alpha = 0.2$ at various
times $ 10^3 < t < 5 \! \times \! 10^4$ (in the dimensionless units of
Eq.\ (\ref{pf_eq})), corresponding to heights ranging from $20 < H <
100$. Although it is difficult to obtain good statistics for this
quantity, immediately apparent is a strong power law decay, $S(k) \sim
1 / k^{1+2 \chi}$, with a global roughness exponent $\chi \approx 1.25$,
and a crossover to a plateau corresponding to distances larger than
the time-dependent correlation length $\xi_{\times} (t)$.

The form of the structure factor indicates the presence of anomalous
scaling; with $\chi > 1$ the surface is superrough \cite{Krug_97} and
the spatial correlation functions will reflect this fact. The level of
$S(k,t)$ in the region of the power decay also seems to increase with
time, i.e.\ $S(k,t) \sim t^{\theta} / k^{1+2 \chi}$, with $\theta
\approx 0.05$, which could indicate the presence of intrinsic
anomalous scaling \cite{Lopez_97}. A clear identification of this
regime is however quite difficult, due to the very slow increase in
time and to poor statistics. 

\begin{figure}
\epsfxsize=3.2in \epsfysize=3.2in
\epsfbox{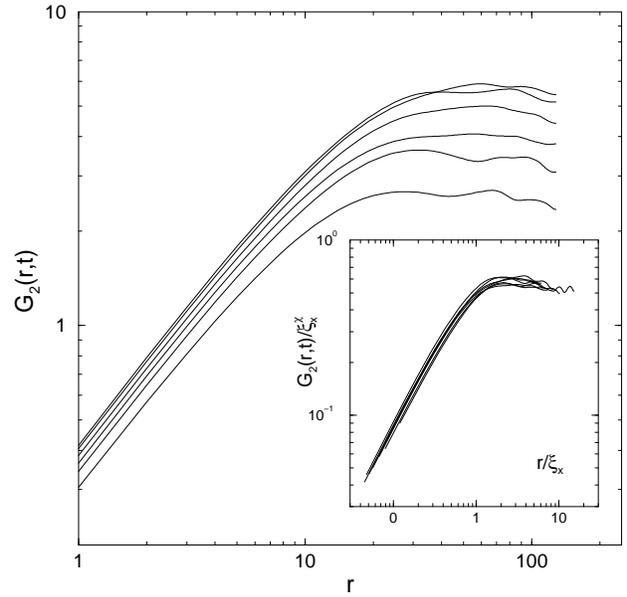}
\caption{Spatial correlation functions $G_2 (r,t)$ for parameters 
identical to those of Fig. \ref{strf_fr}.
The anomalous scaling is indicated by the
increase of the average mean step height $G_2 (r=1,t)$ and the local
exponent $\chi_{loc} \sim 0.9$. In the inset, the data are rescaled
according to Eq.\ (\ref{sc-corr-total}). The scaling shows the
existence of a lateral length scale $\xi \sim t^{1/4}$.
}
\label{G_r}
\end{figure}

The anomalous form of the scaling is most visible in the two-point
spatial correlation function $G_2 (r,t)$ as shown on Fig.\ \ref{G_r}
for the same data as for the structure factor. The correlated
roughness of the interface is visible up to a length scale $r_{\max}
\sim \xi_{\times} (t)$, and the average mean step height $G(r=1,t)
\sim \xi_{\times}^{\chi - \chi_{\rm loc}} \sim t^{(\chi - \chi_{\rm
loc})/4}$ \cite{Krug_97,Lopez_97}.

\begin{figure}
\epsfxsize=3.2in \epsfysize=3.2in
\epsfbox{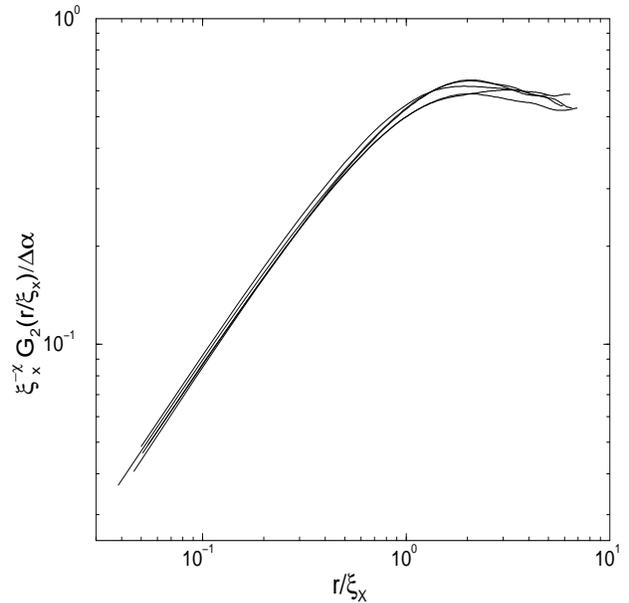}
\caption{Plot of the correlation functions $G_2 (r,t)$ according to
the scaling form Eq.\ (\ref{sc-corr-total}). The data are taken from
systems with lateral size $L = 256$ and gaussian distribution, with
parameters {\bf (i)} $\bar{\alpha} = 0.05$, $\Delta \alpha = 0.2$, at
height $H = 20$, {\bf (ii)} $\bar{\alpha} = 0.1$, $\Delta \alpha =
0.2$, at height $H =  67$, {\bf (iii)} $\bar{\alpha} = 0.2$, $\Delta
\alpha = 0.2$, at height $H = 77$, {\bf (iv)} $\bar{\alpha} = 0.3$,
$\Delta \alpha = 0.2$, at height $H = 102$, {\bf (v)} $\bar{\alpha} =
0.2$, $\Delta \alpha = 0.1$, at height $H = 95$.
}
\label{Xi_alpha}
\end{figure}

The average driving force $\bar{\alpha}$ affects the scaling of the
structure factor and correlation length only through the correlation
length $\xi_{\times} (t) \sim (t/\bar{\alpha})^{1/4}$, and the total
strength of the noise $\Delta \alpha$ only influences the amplitude of
the prefactor of the correlation function. The correlation function
may be fitted to the function
\begin{equation}
G_2 (r,t) = \Delta \alpha \, \xi_{\times}^{\chi} g(r/\xi_{\times}),
\label{sc-corr-total}
\end{equation}
with $\chi = 1.25$ and a scaling function $g(x) = x^{\chi_{\rm loc}} f(x)$
with $f(x) \sim x^{-\chi_{\rm loc}}$ for $x \gg 1$, and approaching a
constant for $x \ll 1$. There seems to be no simple explanation why
$\Delta \alpha$ enters Eq.\ (\ref{sc-corr-total}) in a linear
way
even beyond the linear approximation of Eqs.\ (\ref{eff_local}) and
(\ref{eff_nonlocal}) where it is easy to see.
The {\it local} scaling exponent $\chi_{\rm loc} \simeq 0.9 $ is a
direct consequence of anomalous scaling
\footnote{In principle $\chi_{\rm loc} \! = \! 1$ for a superrough 
interface.  However, here the scaling behavior in $S(k,t)$ appears 
only over a relatively short range and finite size effects are pronounced.}. 
Likewise, the
structure factor may be described by the scaling form
\begin{equation}
S(k,t) = \frac{s(k \xi_{\times})}{k^{1+2\chi}},
\end{equation}
where the scaling function $s (x)$ is constant for $x \gg 1$ and $s(x)
\sim x^{1+2\chi}$ for $ x \ll 1$. The scaling behaviour can be seen
in the inset of Fig.\ \ref{G_r}, where the scaled form of the
correlation function is shown for a single system of $L=256$, with
$\bar{\alpha}=\Delta \alpha =0.2$ at various times and in Fig.\
\ref{Xi_alpha} where the scaled correlation function is now shown for
systems of similar lateral extent but different values of the driving
force and strength of disorder. Within this scaling picture, the early
time development of the width follows 
$W(t) \sim \xi_{\times}^{\chi} \sim t^{\chi/4} \equiv t^{\beta}$  
yielding a growth  exponent $\beta = \chi / 4 \approx 0.31$ in good
agreement with the direct numerical estimate. 

\subsection{Fronts at Fixed Height}
\label{horvath_exp}

The results presented in the last section indicate a scaling picture
to be valid in the freely rising case. It is however difficult to
obtain sufficient statistics and larger samples of quenched disorder 
replicae are necessary to get accurate data. This difficulty can
be overcome by considering the stationary interface, as described by
Eq.\ (\ref{horvath_eq}). In this case, the interface fluctuations
reach a steady state and the various correlation functions can be
obtained with greater accuracy. This setup also allows the
investigation of height difference temporal correlation functions and
permits a comparison with the experimental results of Horv\'ath and
Stanley \cite{Horvath_95}.

Equation (\ref{horvath_eq}) was integrated numerically for different
values of mean height $H = \bar{\alpha}/2 v$ (see Eq.\
(\ref{fixed-h})). Different lateral system sizes were used between $L
\! = \! 12$ and $L \! = \! 400$. The total vertical extent of the
lattice was taken to be about $50$ length units higher than the
interface, to prevent any influence of the upper boundary. The mean
value $\bar{\alpha}=0.2$ in all cases, but different distributions
were used: {\bf (i)} a uniform distribution with mean $\bar{\alpha} =
0.2$, on the range $[0,0.4)$ (standard deviation $\Delta \alpha =
0.07$), {\bf (ii)} the same, but on the range $[0.1,0.3)$, ($\Delta
\alpha = 0.03$ ), {\bf (iii)} an exponential distribution with average
$0.2$ ($\Delta \alpha = 0.2$). Different numbers of configurations were
used in taking the averages, from 10 in the largest systems ($L \! =
\! 2H \! = \! 400$) to 100 in the smallest ($L \! = \! 2H \! = \!
50$). Even in the largest systems, saturation  of the interfacial
fluctuations became apparent after times $t \! \simeq \! 2 \times
10^4$. All systems were integrated up to $t=10^5$, with the interface
configurations extracted at time intervals $\Delta t = 100$.

\subsubsection{Spatial correlations}

\begin{figure}
\epsfxsize=3.2in \epsfysize=3.2in
\epsfbox{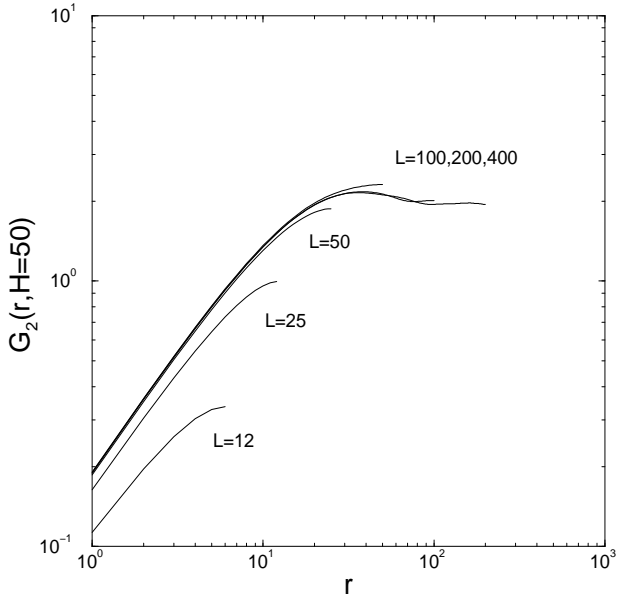}
\caption{Spatial height difference correlation functions for a setup
with fixed average height $H \! = \! 50$, and $\alpha({\bf x})$
uniformly distributed on the range 
$[0,0.4)$. The data for $L \! = \! 100$, $200$ and
$400$ fall together (top curves), while small systems show
$L$ dependence ($L \! = \! 50$,$25$ and $12$, middle to bottom).}
\label{XiL}
\end{figure}

Along with the height $H$, the length scale $\xi_{\times}$ also
remains fixed in this setup, since it is related to the driving
velocity through $\xi_{\times} \sim v^{-1/2} \sim H^{1/2}$. Thus,
contrary to the standard picture of kinetic roughening, the saturation
of the interface is not necessarily determined by the total lateral
extent $L$ of the system. Here, the correlations saturate at {\it
either} the system size $L$ or the correlation length $\xi_{\times}$,
whichever is smaller. Fig.\ \ref{XiL} shows data obtained for $H \! =
\! 50$ on system sizes $L \! = \! 12, 25, 50, 100, 200$ and $400$. The
curves for $L \! = \! 100, 200$ and $400$ collapse indicating that the
fluctuations are bounded by the $L$ independent length scale
$\xi_{\times}$ ($50 \! < \! \xi_{\times} \! < \! 100$ for this
particular case).

\begin{figure}
\epsfxsize=3.2in \epsfysize=3.2in
\epsfbox{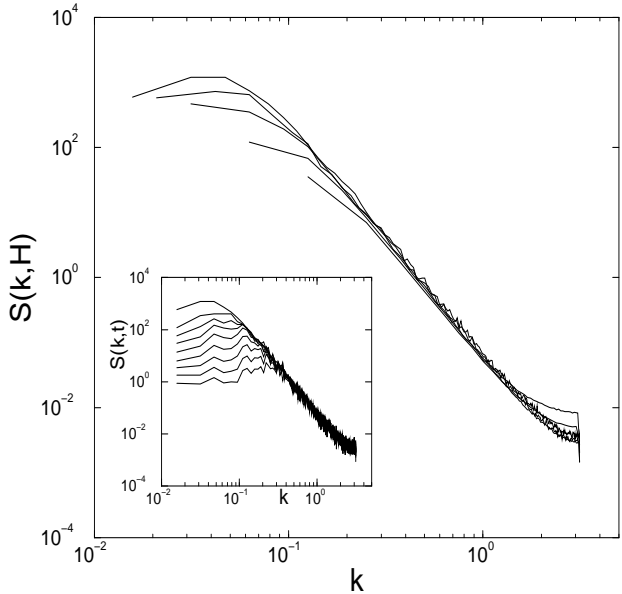}
\caption{Structure factors $S(k)$ for $H \! = \! 25$, $50$, $100$,
$150$, $200$ (bottom to top) and $L \! = \! 2 H$, disorder
chosen from an exponential distribution with $\bar \alpha \! = \!
0.2$. The power law decay is proportional to $k^{-3.5}$, indicating a
global roughness exponent $\chi \! = \! 1.25$. The large scale cutoff
$k^* \! \sim \! 1/\xi_{\times}$ decreases with $H$. The inset shows
the approach of $S(k,t)$ to the saturated $S(k)$ in the system of size
$L \! = \! 2H \! = \! 400$ and for times $t \! = \! 2^n \times 10^2$
with $n=0,1,...7$. 
}
\label{S-k-H}
\end{figure}

The structure factor $S(k,H)$, shown in Fig.\ \ref{S-k-H}, also has a
pronounced power law decay $k^{-(2 \chi + 1)}$ with a global roughness
exponent $\chi \! = \! 1.25$. Again, the interface is
superrough. As in the freely rising case, there
 also seems to be a very weak intrinsic anomaly in
the sense of \cite{Lopez_97}, i.e.\ the prefactor of $S(k,H)$ in
the power law region depends on $H$. The data are consistent with
$H^\theta$ for $0 \! \leq \! \theta \! \leq \! 0.1$, but not accurate
enough to draw any firm conclusion here.

\begin{figure}
\epsfxsize=3.2in \epsfysize=3.2in
\epsfbox{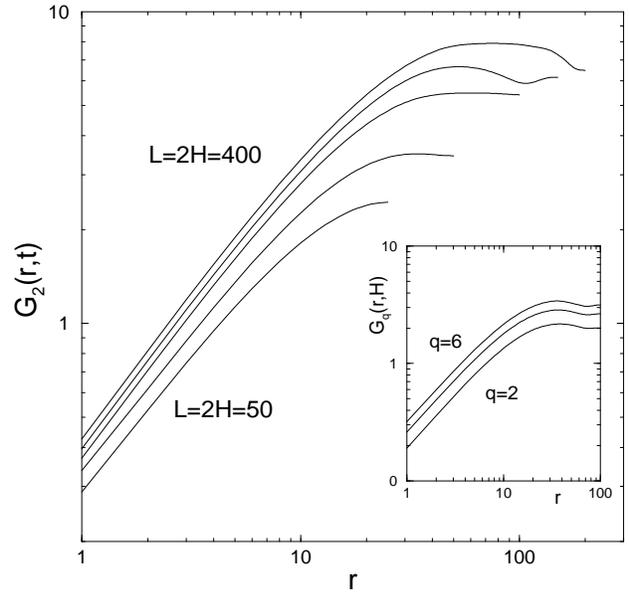}
\caption{Correlation function $G_2(r)$ for the same data as in Fig.\
\protect \ref{S-k-H}.  The curves saturate at a length $\xi_{\times}$
which increases with $H$, together with the ``step height'' $G_2(r \!
= \!1)$ . The local scaling exponent $\chi_{\rm loc} 
\approx 0.95$, close to
the expected value $\chi_{loc} \! = \! 1$. In inset, the higher
moments of the correlation function $G_q (r,H)$ are shown for a system
with $L=200$ and $H=50$. All moments have the same local exponent
$\chi_{loc,q} \sim 1$.  
}
\label{G-r-H}
\end{figure}

\begin{figure}
\epsfxsize=3.2in \epsfysize=3.2in
\epsfbox{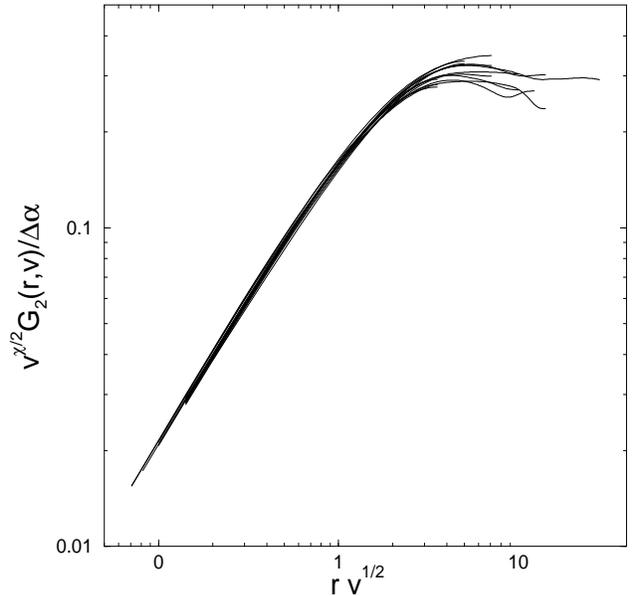}
\caption{Scaling plot of $G_2(r,v)/(\Delta \alpha \; 
\xi_{\times}^\chi)$ vs.\ $r/\xi_{\times}$ 
with global roughness exponent $\chi \! = \! 1.25$
and $\xi_{\times} \sim v^{-1/2} \sim H^{1/2}$. A wide
range of $H$, from $25$ to $400$, and all three forms of disorder are
used. The data compare well to the scaling relations derived in
Eq. (\ref{scaling_g2}).}
\label{Scalplot_g2}
\end{figure}

Provided that $\xi_{\times} \! < \! L$ the spatial correlation
function $G_2(r,H)$ follows the scaling form
\begin{equation}
\label{scaling_g2}
G_2(r,H) = \Delta \alpha \; v^{-\chi/2} \; g(rv^{1/2}).
\end{equation}
with the scaling function $g(x)$ defined as in Eq.\
(\ref{sc-corr-total}). For all different setups, $L \! = \! 50$ to
$400$, $H \! = \! 25$ to $200$, and all three choices for the disorder
$\alpha({\bf x})$, $G_2(r,H)$ is shown rescaled according to Eq.\
(\ref{scaling_g2}) in Fig.\ \ref{Scalplot_g2}. 
For small distances $r
\! < \! \xi_{\times}$ the spatial correlation function is of the form
$G_2(r) \sim r \xi_{\times}^{\chi-1}$, yielding a local roughness
exponent $\chi_{\rm loc} \! \approx \! 1$ and a height difference at
fixed $r$ growing as $\xi_{\times}^{\chi-1}$ \cite{Krug_97} (see
Fig.\ \ref{G-r-H}).

Finally, the different moments $q \! = \! 2,4$ and $6$ of the
correlation functions $G_q (r)$ can be compared, as shown in the inset
of Fig.\ \ref{G-r-H}. All moments have a local exponent $\chi_{loc}
\sim 0.95$. The global exponent $\chi_q$ can in principle be obtained
from the short distance scaling of $G_q (r=1) \sim
\xi_{\times}^{\chi_q-1}$. The present data point to a similar value
$\chi_q = 1.25$ for all $q's$ but are however too noisy to
draw any definite conclusion. Thus, the interface is probably truly
self-affine up to the crossover scale --- be it the system size $L$ or
the saturation length $\xi_{\times}$. 

\subsubsection{Temporal correlations}

Before the interface fluctuations reach the steady state they are
governed by an increasing dynamical correlation length. Starting from
a flat front  $h(x,t\!=\!0) \! \equiv \! H$ it is observed to grow
roughly as $\xi_t \! \sim \! t^{1/3}$ and then approach
$\xi_\times$. However, this behaviour could not be analysed in much
detail, due to the short time range of the initial power law and 
to poorer statistics (averaging over time is not possible). The
insets in Fig.\ \ref{S-k-H} shows $S(k,t)$ approaching the saturated
regime for the large system, $L \! = \! 2H \! = \! 400$.

\begin{figure}
\epsfxsize=3.2in \epsfysize=3.2in
\epsfbox{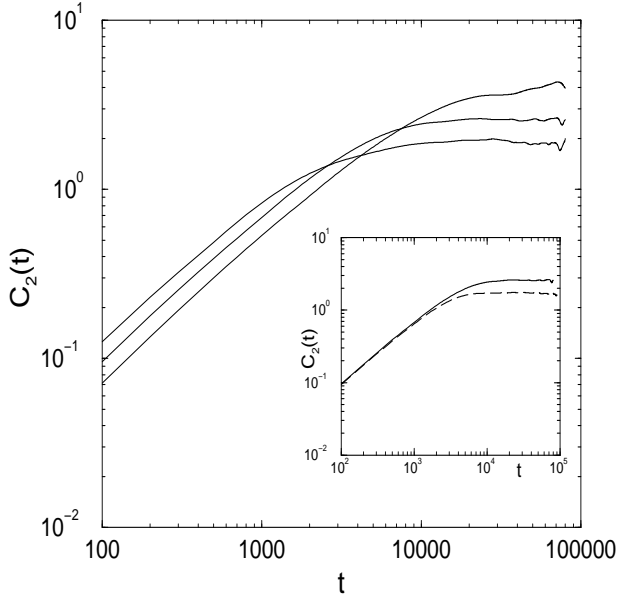}
\caption{Correlation functions $C_2(t)$ for $L \! = \! 400$ and 
$H \! = \! 50,100,200$ (solid curves). The crossover time $t_s$ and the
level of saturation increase with $H$. 
At short times $C_2(t)$ decreases with $H$. 
In inset, $C_2(t)$ for $L \! = \! 50$ and  $H \! = \! 100$
(dashed line) is compared to $C_2 (t)$ for $L \! = \! 400$ and $H \! =
\! 100$ (solid curve). The transition to the saturated regime is
sharper for smaller $L$.}
\label{C_t_H}
\end{figure}

Next, the correlation functions $C_2(t)$, shown in 
Fig. \ref{C_t_H} for different heights are
compared, as in the experiments of Horv\'ath and Stanley
\cite{Horvath_95}. In both cases, the crossover time $t_s$ between the
power law regime of $C_2(t)$ and saturation increases with $H$. This
is true for the level of saturation (or the width $W$) as well. At
early times $t \ll t_s$, the absolute value of $C_2(t)$ {\em
decreases} with $H$.

The data can be related to a scaling function of the form
\begin{equation}
C_2 (t) \sim H^{\chi/2} f (t/H^{z/2}),
\label{scale-C2}
\end{equation}
with the scaling function $f(x) \sim x^{\beta_2}$ for $x \ll 1$ and
{\it const.} for $x \gg 1$ with a (genuine) dynamical exponent
$z \approx 2$ and the effective slope $\beta_2 \approx 0.85$. 
Although the exact
value of these exponents is difficult to establish,
this form is however valid for all $L$ provided that
$\xi_{\times} < L$.

\begin{figure}
\epsfxsize=3.2in \epsfysize=3.2in
\epsfbox{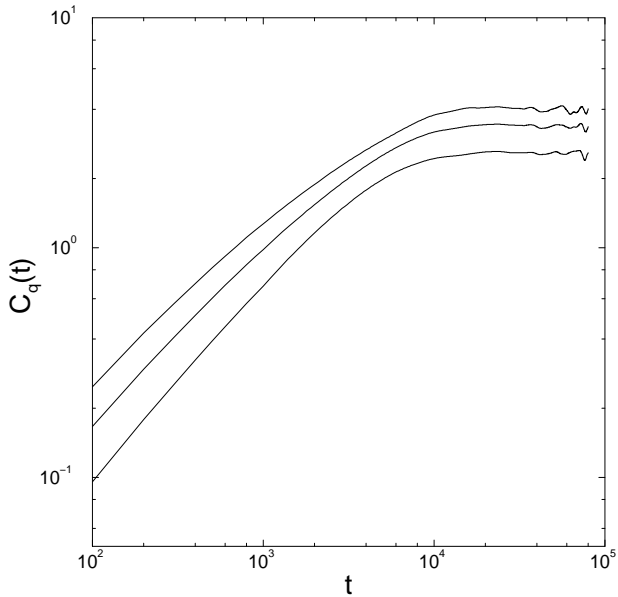}
\caption{
Correlation functions $C_q(t)$ for $q \! = \! 2,4$ and $6$ for systems
of size $L \! = \! 200$ and $H \! = \! 50$, such that  $L \! > \!
\xi_{\times}$. Each moment increases with a {\em different} exponent
$\beta_q$.
}
\label{temporal_scaling}
\end{figure}

The different moments $C_q(t)$ for $q \! = \! 2,4$ and $6$ are
shown in Fig.\ \ref{temporal_scaling}. They clearly have different
behaviour. The early time logarithmic slopes of the higher moments
($\beta_q$) {\em decrease} with $q$, as shown in the inset of Fig.\
\ref{temporal_scaling}. For the higher moments, the effective
exponents $\beta_4 \approx 0.76$ and $\beta_6 \approx 0.69$. Such
multiscaling has been observed in cases connected with the existence
of {\it avalanches} in the interface dynamics \cite{LeschhornTang}. It
is clear from Fig.\ \ref{fronts_fig} that similar avalanche type of
motion exists here as well, but only up to the vertical length scale
$W$ and lateral size $\xi_{\times}$. Because of this reason the
quantitative characterization of such avalanches is beyond the scope
of the present work.


\section{Discussion}
\label{analysis}

\subsection{Temporal Scaling of the Interface and Relation
to Experiments}

\begin{figure}
\epsfxsize=3.2in \epsfysize=3.2in
\epsfbox{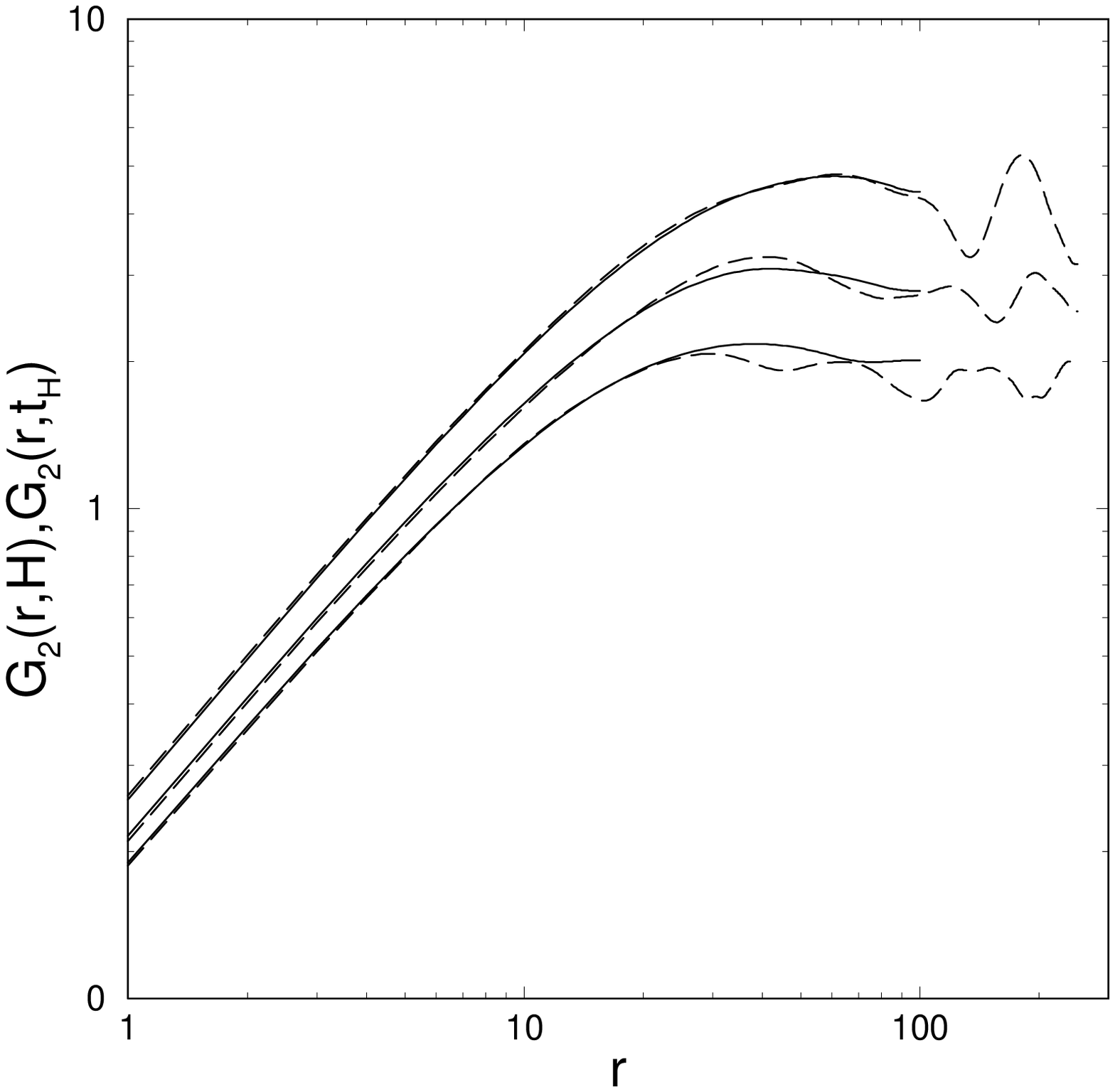}
\caption{Comparison of the correlation function $G_2(r,H)$ in the
stationary setup (solid lines) at heights $H \! = \!  25, 50$ and
$100$ to  $G_2(r,t_H)$ in the freely rising case (dashed lines) at
corresponding times $t_H=H^2/\bar{\alpha}$. There is a complete
equivalence between both situations.
}
\label{quasistation}
\end{figure}

It is interesting to first compare the results for the stationary
fronts with those of the freely rising fronts. Both cases are governed
by the same height dependent length scale $\xi_{\times}$. In Fig.\
\ref{quasistation} the spatial correlation function $G_2(r,H)$ (fixed
$H$ at saturation for the stationary case) and $G_2(r,t_H)$ (freely
rising case at times $t_H = H^2/ \bar{\alpha}$ when the average height
has reached $H$) are shown for various values of $H$. There is a
complete equivalence between the interfacial fluctuations at an
instantaneous height $H(t)$ and the saturated fluctuations of a
stationary interface. In {\em both} cases, the range of correlated
roughness is determined by the same value $\xi_{\times}$. The length
scale $\xi_\times$ is thus conceptually different from the intrinsic
time dependent lateral correlation length $\xi_t$ commonly found in
models of kinetic roughening. Here $\xi_{\times}$ merely fixes the
maximum range of correlated roughness. Such ``quasi-stationarity'' of
the moving front can only occur provided that the ``natural''
dynamical exponent $z < 4$, so that the interface fluctuations can
always catch up instantly with the available area of correlation.

The model can also help to interpret the experiments performed by
Horv\a'ath and Stanley on the stationary interface \cite{Horvath_95},
since it yields qualitatively similar results: The exponent $\beta_2$
is constant for all driving forces (and also $\beta_q$ for the higher
moments considered), the level of saturation of $C_2(t)$ increases as
$H$ increases, while the amplitude of the early time power law
behaviour decreases.

Still, there is a quantitative difference between the experiments of
\cite{Horvath_95} and the numerical results. In the present work,
$\beta_2 \! = \! 0.85$ in contrast to the experimental value
0.56. Clearly one would not expect coincidence, since already the
average front velocity behaves differently: The model shows ``pure''
Washburn behaviour, $dH/dt \! \sim \! 1/H$, whereas in
\cite{Horvath_95} $dH/dt\! \sim \! H^{-1.6}$. This deviation from
Washburn's law occurs in many paper imbibition experiments
\cite{Dube_Exp}, and there is no reason why its origin should not
affect the intrinsic fluctuation dynamics as well, leading to
different values of $\beta_q$ in the time correlation functions.

There is however a clear contradiction of the results with the scaling
form for the time-correlation function suggested in \cite{Horvath_95},
\begin{equation}
\label{c-t-horvath}
C_2 (t) \sim V^{-\theta_L} L^{\chi} \; {\mathcal C} (t L^{-\chi/\beta}
V^{(\theta_t + \theta_L)/\beta})
\end{equation}
with the relation $V = dH/dt \sim H^{-1.6}$ and the scaling function
${\mathcal C} (u)$ such that $C_2 (t) \sim t^{\beta} V^{\theta_t}$ if
$u \ll 1$ and $C_2 (t) \sim L^{\chi} V^{-\theta_L}$ in the opposite
limit. 
Equation (\ref{c-t-horvath}) does not include an explicit lateral length
scale $\xi_\times$, although some power of $V^{\theta_t + \theta_L}$ 
may obviously play such a role. This scaling form is however in
contradiction with the present results in two ways. First, provided
that $\xi_{\times} > L$, the total lateral system size does not play
any role in the scaling of the interface. Secondly, no common scaling
form interpolating from large systems (fluctuations up to scale
$\xi_{\times}$) to small systems (dominated by $L$) has been found. The
saturation of $C_2(t)$ is sharper for small $L$, which is visible in
Fig.\ \ref{C_t_H}.

In the experiments only one lateral system size was used, so the
role of the system size $L$ could not be assessed. Unfortunately, no
information on the spatial scaling of fluctuations was presented so
that no comparison can be done to the present work. Likewise, the
higher moments of the temporal correlation function were not measured
and the presence of avalanches could not be inferred.

\subsection{Avalanches, Pinning and Roughness of the Interface}

As one can see from Figure~\ref{fronts_fig} the interface motion
indicates the presence of ``avalanches'' as in usual models 
describing depinning transitions \cite{Kardar_98}. Here the behavior
is however somewhat different since the flow of liquid in the 
average tends towards regions of lower chemical potential. In some way,
this is analogous to 'self-organised' interface models in which the
interface is driven at the point where the force is the largest 
\cite{LeschhornTang,Sneppen_92,Olami_94}.

The question is now whether it is possible to understand the observed
exponents --- $\chi, \beta_q$ etc. --- in terms of an avalanche
description. Were this to be true, the presence of the conservation
law would only be felt through the correlation length $\xi_\times$,
which would limit the avalanche area to $\xi_{\times}^{\chi+1}$ with an
exponential cut-off on the avalanche size distribution 
(see \cite{Dougherty_98} for a discussion of a
related experiment). It turns out that the multiscaling observed here
($\beta_q$) differs in a crucial way from that obtained for avalanches
in self-organised depinning. Leschhorn and Tang \cite{LeschhornTang}
obtain an almost trivial multifractal spectrum for the $\beta_q$'s
assuming that a local dynamical exponent $z_{\rm loc}$ and a global
roughness exponent can be defined. Dynamical scaling in this sense is
however absent in the present case, since the 
interface dynamics depends directly on the local height.

The observed global roughness exponent $\chi = 1.25$ also appears in
``nonconserved'' front propagation through a medium with quenched
disorder \cite{Leschhorn_93,Roux_94,Jost_96}, i.e.\ for interfaces in
the quenched Edwards-Wilkinson (QEW) universality class. In that
problem a similar value of $\chi$ is observed close to criticality, 
whereas a cross--over to thermal EW (the massless Gaussian field, 
with $\chi = 0.5$) takes place in the moving phase. 
The imbibition model may then present a similar behaviour due to the 
continuous slowing down of the interface,
with parts of the interface pinned (thus approaching the depinning
transition from above).  As in the QEW class, the value 
$\chi = 1.25$ means that the width increases faster than the 
correlation length. The ``local slopes'', 
$G_2 (r=1,t) \sim \xi_{\times}^{\chi-\chi_{\rm loc}} (t)$ 
diverge with time and the local roughness exponent $\chi_{loc} \approx
0.9 \dots 0.95$. For QEW models this is true on any scale --- up to
saturation in a finite system --- but in our case the behavior can
hold only up to $W \simeq \xi_{\times}$, because the removal of overhangs
occur naturally in a phase field formulation. 
If $W > \xi_{\times}$ overhangs in neighboring bulges of the 
interface can merge, always keeping $\xi_{\times}$ of the order of $W$.

A possible mechanism by which the roughness might change is if
the interface becomes completely pinned. In Appendix \ref{real}, 
it is shown
that this may be achieved through the inclusion of either evaporation
or gravity. In such a case, new length scales come into play, and the
roughness exponent is most likely changed.

Although the pinning effects of evaporation have not been explicitly
considered in this work, preliminary calculations show that
evaporation, characterised by an evaporation rate $\epsilon$
introduced in the Appendix \ref{real}, pins the interface at a height 
$H_p \sim (\bar{\alpha}/\epsilon)^{1/2}$. At this pinning height, a
new correlation length $\xi_{\epsilon}$ emerges. For weak evaporation,
defined as $\epsilon \ll \bar{\alpha}^3/\sigma^2$, the length
$\xi_{\epsilon} \sim (\sigma/\epsilon)^{1/3} \gg
\xi_{\times}(H_p)$. For strong evaporation, $\epsilon \gg
\bar{\alpha}^3/\sigma^2$, this length is $\xi_{\times} \sim
(\sigma/\epsilon H_p)^{1/2} \sim \xi_{\times} (H_p)$. This question,
as well as the appearance of $\xi_\times$ itself, has not been addressed
by previous models and experiments \cite{Dube_Exp}.

Within this region, there should be a crossover from the superrough
interface with $\chi =1.25$ to a pinning regime. The roughness
exponent on scales below the correlation length should then 
be determined solely by the local disorder configuration, i.e. be related 
to e.g. directed percolation depinning.

\subsection{Hydrodynamical description of Imbibition}

The model presented in Section \ref{modelpresent} is purely diffusive
and does not include any hydrodynamical modes. In principle, these
could be incorporated, in a coarse-grained sense, along the lines of
Refs.\ \cite{Kawasaki_82} and \cite{Jasnow_96}
by coupling the phase field to a velocity field described by
the Navier-Stokes equations. 
Unfortunately, many
problems need to be resolved before such an approach is taken, the
main one consisting in establishing the role of hydrodynamics
itself \cite{Dube_Exp}. Contrary to other porous media
like fractured rock or Hele-Shaw cell filled with glass beads,
the paper matrix used in most imbibition experiments is not inert 
but often interacts strongly with the invading fluid through fiber swelling. 
Another complication arises from
the transport of fluid through the paper. It is not at all obvious that
an homogeneous pressure can be defined throughout the volume
occupied by the fluid. In that sense, the model defined by Eq.\
(\ref{pf_eq}) and Eq.\ (\ref{horvath_eq}) is the minimal model that
includes a conservation law and, in conjunction with the
appropriate set of boundary conditions, reproduces the experimental
characteristic of imbibition.

Even though hydrodynamics is absent 
there is however of course a strong similarity between the 
model and the standard
description of flow in porous media based on Darcy's Law for an
incompressible fluid. In this description, the normal velocity of the
interface is related to the gradient in the pressure field $P({\bf
x})$ by the permeability $\kappa$ as $v_n = - \kappa \; \partial_n
P(x,y \!\! = \!\! h(x,t))$. The pressure is determined from Laplace's
equation, together with $P(x,y=0) = P_0$, the atmospheric pressure,
and a Gibbs-Thomson boundary condition at the
interface: $P(x,h) = P_0 - P_c (x,h(x,t)) -\gamma \nabla^2 h(x,t)$, 
which introduces a coarsed-grained surface tension \cite{Krug_91} and
capillary pressure $P_c$ arising from the microscopic menisci at the
fluid--gas interface. Working to linear order in the small
fluctuations of the interface, it is straightforward to find the
pressure field, defined for $y \leq h(x,t)$, 
\begin{equation}
P(x,y)  = P_0 - \bar{P_c} \frac{y}{H} + \sum_k e^{ikx}
\sinh(ky) \,P_k \, ,
\end{equation}
with the coefficient
\begin{equation}
P_k = \frac{1}{\sinh (kH)}
\left( \left( \gamma k^{2} - \frac{\bar{P_c}}{H} \right) h_k
- P_c (k \neq 0,H) \right) \, ,
\end{equation}
where $\bar{P_c}$ represents the average capillary pressure and 
$P_c(k \neq 0,H)$ the fluctuations around it. It is then
a simple matter to derive the interface equations, Eq.\
(\ref{eff_local}) and Eq.\ (\ref{eff_nonlocal}) in the appropriate
limits. The derivation of the length scale $\xi_{\times}$ may be
transposed directly to the fluid imbibition case. This is essentially
equivalent to the approach of Ref.\ \cite{Ganesan_98} (see also
\cite{Zik_97}), although spontaneous imbibition requires a special
treatment of the boundary conditions, absent in their work. 

Our model is based on a constant mobility, which in the general case
should be replaced by $\partial_t \phi = \nabla M ({\bf x}) \nabla
\mu({\bf x},t)$. In spite of its simpleness, it is quite reasonable to 
ask whether the quenched randomness should not be included in the mobility, 
to model an effective quenched permeability. In most cases of {\it forced}
fluid flow in bulk random media, this is where the non-uniformities
are most relevant \cite{Koponen_98}.
One point must however be emphasised. 
In any imbibition experiment designed for this purpose, the flow will
never be large, and we believe that the random capillary forces will
have the dominant influence. 
On the other hand,
it has recently been shown that the presence of ink, or presumably of
any other blocking material does make a quenched porosity relevant
\cite{Zik_98}, a case which is not considered here.

It should also be pointed out that the field $\phi$ does not represent
a real fluid density. In particular, any ``air bubbles'' (i.e.\
connected regions of value $\phi = -1$) trapped behind the front will
eventually dissolve. This is of course highly unrealistic for bulk
porous media, but may be appropriate for thin porous media, where air
can escape through surface pores.


\section{Conclusion}

In conclusion, a simple phase field model for the invasion of a liquid
into a disordered system has been introduced. Liquid conservation is
explicitly included. Some basic features observed in imbibition
experiments are reproduced. Of course, it cannot account for many
phenomena of paper wetting or invasion into porous media, which are
briefly discussed in Section \ref{analysis} and treated in more detail
in the previous paper \cite{Dube_Exp}.

In numerical simulations, a superrough interface is found, with
anomalous scaling due to a global roughness exponent $\chi \simeq 1.25
> 1$. A hardly discernible inherent anomaly of the structure factor
may also be present. The extent of the spatial fluctuations of the
interface are bounded by a length scale $\xi_{\times} \sim ( \sigma H
/\bar{\alpha})^{1/2}$, both in the freely rising and stationary
imbibition setup. Interface fluctuations of a rising front are
quasistationary, in the sense that at any time $t$ they are the same
as in a stationary front kept at average height $H \! = \! H(t)$. The
temporal fluctuations show multiscaling, which indicates motion by
avalanches. The length and time scales where fluctuations saturate can
be understood by simple dimensional considerations.

In relation to experiments, according to this analysis it is highly
desirable to have fluids with small capillary pressure and high
surface tension in order to obtain scaling over a large spatial
regime.  It
would also seem appropriate to use organic liquids in experiments done
with paper. These have minimal chemical interaction with the
constituent fibers, and a simplified Washburn description of
imbibition may be applicable. Another option is to use deionised water
\cite{Zik_98}, again with the goal of reducing the fiber-liquid
interaction. Above all, the main conclusion of the present work is
that the macroscopic behaviour of the average interface position
(i.e.\ $H(t)$ in a freely rising study, or $H(v)$ if the interface is
stationary) is crucial to an understanding of the microscopic
fluctuations of the interface, since it controls the range over which
scaling can be observed.

\section*{Acknowledgements}
We are grateful for discussions with and comments from R.\ Cuerno, R.\
Hilfer, V.\ Horv\'ath, V.\ Kiiski, J.\ Krug, E.\ Sepp\"al\"a
and M.\ Smock. This work was supported by the Academy of Finland and
the Research Corporation under grant No. CC4787 (KRE). S.\ M.\ 
acknowledges travel grant from the Magnus Ehrnrooth Foundation.

\section*{Appendix}
\appendix

\section{Refining the Phase Field Model}
\label{real}

In this appendix, a generalised imbibition model is introduced.  It
rests on the following dynamical equation for the dimensionfull 
phase field $\tilde{\phi} ({\bf r},t)$
\begin{equation}
\frac{\partial  \tilde{\phi} ({\bf r},\tau)}{d \tau} - \tilde{G} 
\frac{\partial \tilde{\phi} ({\bf r},t)}{\partial y} = 
\nabla M(\tilde{\phi}) \nabla
\frac{\delta F}{\delta \tilde{\phi}} - 
\tilde{\epsilon} (\phi_e+\tilde{\phi}({\bf r},\tau))
\end{equation}
with the free energy functional
\begin{equation}
F = \frac{1}{2} \int d{\bf r}  \left[ r \tilde{\phi}^2
+ \frac{u}{2} \tilde{\phi}^4 + \kappa (\nabla \tilde{\phi})^2 
 - \tilde{\alpha} \tilde{\phi} \right] \, ,
\end{equation}
and $\phi_e = (r/u)^{1/2}$. 
For $\tilde{G}=0$ and $\tilde{\epsilon}=0$, 
as well as for constant mobility 
$M(\tilde{\phi}) =M$, this reduces to the model described in the
introduction. When $\epsilon \neq 0$, a non-conserving term is
introduced in the equation of motion, which, in a first
approximation describes evaporation of liquid (the $\tilde{\phi} = 
\phi_e$ phase), at a rate $2 \phi_2 \epsilon$ and proportional 
to the total area 
covered by the fluid. The convective term is included to describe
gravity. Although gravity can only be introduced properly through an
hydrodynamical field \cite{Kawasaki_82,Jasnow_96}, it is shown below
that this term reproduces the correct equation of motion for the
average position of the imbibition front.

Assuming a constant mobility $M$, these equations can be 
put in a dimensionless form by defining 
\begin{equation}
\begin{array}{ll}
{\bf x}  =  {\bf r}/\zeta  \, ; 
\hspace{2.0cm} & 
\alpha = \left[ \frac{u}{r^3} \right]^{1/2} \tilde{\alpha}  \, ; \\
t = \left[ \frac{ M r^2}{\kappa} \right] \tau \, ;  & 
G = \left[  \frac{\kappa}{r^3}\right]^{1/2}  \frac{\tilde{G}}{M} \, ;\\ 
\phi = \frac{\tilde{\phi}}{\phi_e} \, ; & 
\epsilon  = \frac{\kappa}{M r^2} \tilde{\epsilon} \, . 
\end{array}
\end{equation}
The ratio $\zeta = (\kappa/r)^{1/2}$ determines the width of
the interfaces between different phases.

For now, let us concentrate on the case $G =\epsilon = 0$. 
In one dimension, with $\bar{\alpha}$ being constant, 
the chemical potential obeys a
Laplace equation in both wet and dry phases (actually there are minor
corrections) with boundary conditions, $\mu(y=0)=0$ and $\mu(y=H)$ $=$
$\mu(y=L_y)$ $=$ $-\bar{\alpha}$, where $H$ is the position of the
wet/dry interface and $L_y$ is the length of the paper.  Thus
\begin{equation}
\mu (y) = \left\{ \begin{array}{ll}
        -\bar{\alpha} y/H, \hspace{1.0cm} & \mbox{if $y \leq H$}; \\
         -\bar{\alpha},  & \mbox{if $y > H$};
        \end{array}
\right.
\end{equation}
which implies that $\phi$;
\begin{equation} \label{pf_corr}
     \phi(y) = \left\{ \begin{array}{ll}
      \phi_o+(1-\phi_o) \; y/H, \hspace{0.5cm} & \mbox{if $y \leq H$}; \\
       -1,   & \mbox{if $y > H$};
        \end{array}
\right.
\end{equation}
where $\phi_0 > 1$ is the solution of $-\phi+\phi^3 = \bar{\alpha}$.
In a first approximation, both $\phi$ and $\mu$ are linear functions 
of $y$ for $y\leq H$. The total amount of concentration is then
\begin{equation}
\Phi_{tot}(t)= \int_{0}^{L_y} \phi(y,t) dy = 
\frac{1}{2}(\phi_o+3)H(t) - L_y,
\end{equation}
and the equation of motion for $\Phi_{tot}$ is 
\begin{equation}
\frac{d\Phi_{tot}}{dt} = \frac{\phi_o+3}{2} \frac{dH}{dt} \\
=\int_{0}^{L_y}
\frac{\partial^2 \mu}{\partial y^2} dy , 
\end{equation}
or
\begin{equation}
\frac{d H(t)}{d t} = 
\frac{2}{3+\phi_0}
\left(\frac{\bar{\alpha}}{H(t)} \right),
\label{wash-pure}
\end{equation}
an equation similar to Washburn's result. Since $\phi_0 \sim 1 +
O(\bar{\alpha})$, this is actually Eq.\ (\ref{washburn}) of Section
\ref{modelpresent}. In presence of gravity and/or evaporation, the
solution is more involved. As a first approximation, the difference
between $\phi_0$ and unity is neglected and the Poisson equation for
the chemical potential is considered,
\begin{equation}
\frac{ d^2 \mu }{dy^2} - \epsilon = 0 \, , 
\label{mu-poisson}
\end{equation}
for $y \leq H(t)$ and boundary conditions $\mu(y=0)=0$ and $\mu
(y=H(t))=-\bar{\alpha}$. Again, $\mu (y \! > \! H(t)) =
-\bar{\alpha}$. The solution of Eq.\ (\ref{mu-poisson}) is
\begin{equation}
\mu (y,H(t)) = - \frac{\bar{\alpha} y}{H(t)} 
+  \frac{1}{2} \epsilon y (y -H(t))
\label{mu-model-evap}
\end{equation}
Using the same procedure as above, the equation of motion of the
interface is found to be
\begin{equation}
\frac{d H(t)}{d t} =  \frac{\bar{\alpha}}{2H(t)} 
 - G - \frac{1}{4} \epsilon H(t) \, .
\label{wash-mod}
\end{equation}
The ``gravity'' term acts exactly as in Washburn's equation in
presence of gravity, and thus allows to identify $G$ as an effective
gravity force acting on the interface. As far as we are aware, no
detailed studies of fluid propagation in a thin porous  medium with
evaporation has been done. 

Non-zero values of $G$ or $\epsilon$ will eventually stop the
interface at an equilibrium height $H_{eq} = \bar{\alpha} / G$ if
$G \gg \epsilon$ or $H_{eq} = (\alpha / \epsilon )^{1/2}$ if
$\epsilon \gg G$. There is however a conceptual difference between
pinning due to gravity or evaporation. In the former case, the
chemical potential is a linear function of position, and pinning is
determined by $\partial_n \mu (H_{eq}) = G$ while, in the latter
case, the chemical potential is quadratic in $y$ and at pinning,
$\partial_n \mu (H_{eq}) = 0$. When both evaporation and pinning are
present, the equilibrium height is determined by the zero of
Eq.\ (\ref{wash-mod}). 

In terms of a dimensionfull interface height $\tilde{H}$, 
Eq. (\ref{wash-mod}) becomes
\begin{equation}
\frac{d \tilde{H}(\tau)}{d \tau} =  
\frac{1}{2} \frac{M \tilde{\alpha}}{\phi_e} \frac{1}{\tilde{H}(\tau)}
 - \tilde{G} - \frac{1}{4} \tilde{\epsilon} \tilde{H}(\tau),
\label{wash-full}
\end{equation}
where $\tilde{\alpha}$ represents the average value of the disorder.
The motion of the average interface in the pure Washburn case is 
thus determined by a combination of the mobility $M$ and the shift
in the average chemical potential $\tilde{\alpha}$. The length scale
$\xi_{\times} \sim (\tilde{\sigma} \tilde{H} /\tilde{\alpha})^{1/2}$
where now, $\sigma = (2 \sqrt{2} /3 ) (\kappa r \phi_e^4 )^{1/2}$ is
the dimensionfull surface tension. 

\section{Projection to an Interface Equation}
\label{proj}

To extract the interface equation Eq. (\ref{non-local}) in the limit
$G =0$ and $\epsilon=0$, the dynamical phase field 
equation must first be inverted with the use of the Green's
function defined by Eq. (\ref{gr_def}) and Eq. (\ref{gr_form}): 
\begin{equation}
\label{pf_invert}
\int d{\bf x}' \; G ({\bf x}|{\bf x}') \;
\frac{\partial \phi({\bf x}',t)}{\partial t} = 
\mu ({\bf x},t).
\end{equation}
It is then convenient to use a local coordinate system 
$(u,s)$ \cite{VanSaarloos_98}. The 2-dimensional space is spanned by the
the vector ${\bf x}(u,s) = {\bf X}(s) + u {\bf \hat{n}} (s)$, 
where ${\bf X}(s)$ is a point of the interface, ${\bf \hat{n}}$ is a 
unit vector normal to the interface and $s$ is the arc-length 
coordinate. 
In terms of the phase field, this corresponds to 
$\phi(u=0,s)=0$. The time derivative of
the field then becomes $\partial \phi (u,s,t) /\partial t = V_n (s) 
\partial \phi / \partial u$ where $V_n (s)$ is the normal velocity of the
interface at position $s$. If the interface (of
thickness $\zeta = 1$ in dimensionless units) 
is much smaller than the typical radii of curvature
of the interface (the sharp interface limit), 
the Laplacian term of the chemical potential
may be expanded such that 
\begin{equation}
\nabla^2 = \frac{\partial^2}{\partial u^2} + \frac{\partial^2}{\partial s^2}
+ {\cal K} (s) \frac{\partial }{\partial u} 
\end{equation}
where ${\cal K} (s)$ is the curvature of the interface. For 
$\alpha=0$, the one-dimensional kink solution
is $\phi (u,s) = \phi_0 (u) = \tanh (u /\sqrt{2})$. 
The corrections to this form, 
represented by Eq. (\ref{pf_corr}) are of order 
$\zeta \bar{\alpha}/H$. 
To first order thus, $\mu \sim -\alpha (u,s)  - 
{\cal K} (s) \partial \phi_0 (u) / \partial u$. Still in the sharp
interface limit $\zeta {\cal K} \ll 1$, the derivatives of the kink
solutions have properties $\ \partial \phi_0 (u) / \partial u \sim 
\Delta \phi \delta (u)$ and 
$\sigma = \int du ( \partial \phi_0 (u) / \partial u)^2$ where 
$\Delta \phi \sim 2$ is the miscibility gap and $\sigma$ is the 
interface tension. Multiplying  Eq. (\ref{pf_invert}) by 
$\int du  (\partial \phi_0 (u) / \partial u)$ then effectively project
the phase field dynamics onto the interface $u=0$. A translation
$u \rightarrow u + h(s,t)$ then yields
\begin{equation}
\int ds' \; G (s,h(s,t)|s',h(s',t)) V_n (s') \;
= \eta (x,h(x,t)) + \sigma {\cal K}.
\end{equation}
Equation (\ref{non-local}) is then obtained by a further change of
coordinate $s \rightarrow x$ and the relation
$ds' V_n (s') = dx' \partial_t h(x',t)$.

\newpage

\begin{thebibliography}{99}
\bibitem{remark_poor_scaling} It should be noted that the scaling regime of
Ref.\ \cite{Buldyrev_92} is confined to a length scale $l < 1 \,
\mbox{cm}$, i.e.\ a length scale of the order of the mean fiber length
itself. Ref.\ \cite{Amaral_94} addresses this problem indirectly, but
the ``universality'' of the scaling on such short length scales
remains questionable.
\end{thebibliography}
\end{document}